\RequirePackage{arydshln}
\documentclass[floatfix,aps,twocolumn,nofootinbib,superscriptaddress,preprintnumbers,pra,10pt]{revtex4-2}

\usepackage[utf8]{inputenc}
\usepackage{float}
\usepackage{colortbl}
\usepackage{amsmath,amssymb,amsfonts,bm,bbm,slashed}
\usepackage{dsfont} 
\usepackage{hyperref}
\usepackage{graphicx}
\usepackage{enumitem}
\usepackage{mathtools}
\usepackage{bbold}
\usepackage{braket}
\usepackage{multirow}
\usepackage{xcolor}

\usepackage{lipsum}  

\usepackage{pifont}% http://ctan.org/pkg/pifont

\makeatletter
\g@addto@macro\bfseries{\boldmath}
\makeatother

\newcommand{\gsim}{\lower.7ex\hbox{$\;\stackrel{\textstyle>}{\sim}\;$}}
\newcommand{\lsim}{\lower.7ex\hbox{$\;\stackrel{\textstyle<}{\sim}\;$}}

\newcommand{\bea}{\begin{eqnarray}}
\newcommand{\eea}{\end{eqnarray}}

\begin{document}

\preprint{ }

\title{Revisiting mono-\texorpdfstring{$\tau$}{tau} tails at the LHC}
 
\author{Florentin Jaffredo}
\email{florentin.jaffredo@ijclab.in2p3.fr}
\affiliation{IJCLab, Pôle Théorie (Bât.~210), CNRS/IN2P3 et Université Paris-Saclay,
91405 Orsay, France}

\begin{abstract}
\vspace{5mm}
We revisit the constraints on semileptonic operators derived from the mono-$\tau$ high-$p_T$ events in $pp$ collisions at LHC. Like in previous studies, we obtain limits on the New Physics couplings from the CMS data much stronger than from the ATLAS data, despite a nearly equal integrated luminosity. We find that a neglected systematics on the $\tau$ lepton reconstruction efficiency can partly explain that difference. We then provide new limits on the New Physics couplings relevant to decays based on $b\to c\tau\bar{\nu}$ by using the recent ATLAS data with $139\:{\rm fb}^{-1}$. We also show that the inclusion of propagation of the New Physics particle can significantly worsen the bounds obtained by relying on Effective Field Theoretical treatment of New Physics. 
\end{abstract}

\maketitle

\section{Introduction}

To explain the hints of lepton flavor universality violation observed at LHCb and $B$-factories, one needs to go Beyond the Standard Model (BSM). These hints have been observed in the neutral processes based on the $b\to s\ell\ell$ transition with a combined significance of $4.2\:\sigma$ \cite{LHCb:2021trn,Bordone:2016gaq}, as well as in the charged semileptonic $b\to c\ell\bar{\nu}$ decays, with a significance of about $4\:\sigma$ \cite{HFLAV:2019otj,Na:2015kha,MILC:2015uhg,FermilabLattice:2021cdg}.

Semileptonic transitions can also be probed in the LHC, by colliding quarks from protons ($q=u,d,s,c,b$), namely by looking at the tail of $pp\to\ell\ell$ or $pp\to\ell\nu$ as measured by ATLAS and CMS. In other words, one translates the specific searches at the LHC into constraints on New Physics (NP) couplings relevant to low energy processes. It was shown that despite the suppression coming from parton distribution functions (PDFs) of heavy quarks the constraints on NP couplings can be competitive with those obtained by using low energy flavor observables due to the energy enhancement of the partonic cross-section \cite{Eboli:1987vb,Angelescu:2020uug,Fuentes-Martin:2020lea,Greljo:2018tzh,Marzocca:2020ueu,Iguro:2020keo}.

In this letter, we rely on the results of the very recent searches of a $W'$-resonance with a single $\tau$ lepton in the final state based on a data sample of $139 \:{\rm fb}^{-1}$ by ATLAS in Refs.~\cite{ATLAS:2021bjk}, in order to derive the constraints on NP couplings. In addition to providing an update to the previous studies we also examine some common problems of this approach which might lead to overly tight constraints. In that regard, we identified two different issues:

\begin{enumerate}[label=\roman*.]
\item The systematic uncertainty on the $\tau$ lepton reconstruction efficiency reported in the experimental papers was not systematically taken into account in most theoretical reinterpretation of the ATLAS and CMS searches. Since this uncertainty is dominant for $\tau$'s at high energies, where the NP is the most enhanced compared to the Standard Model (SM), this can lead to a bound too strong by about $10~\%$ for ATLAS and $40~\%$ for CMS. This issue is detailed in section \ref{sec:eff}.\\
\item The Effective Field Theory (EFT) approach is not always valid for those searches due to the insufficient separation between the energy scale of the partonic collision and the EFT cutoff $\Lambda$. Naively, by using the model-independent limits in models with mediators lighter than $\Lambda$, even in the case of non-resonant interactions, the resulting bounds are much stronger than those obtained with the propagation of mediators. In section \ref{sec:eft} we illustrate that effect for a scalar leptoquark with a mass of the order of $1\:{\rm TeV}$, and show it can be as large as $60~\%$.
\end{enumerate}

In this work we focus on the charged $b\bar{c}\to \tau\bar{\nu}$. Previous bounds on these processes by using the LHC data with $36\:{\rm fb}^{-1}$, have been reported in Refs.~\cite{Greljo:2018tzh,Marzocca:2020ueu,Iguro:2020keo}. Here, we revisit these bounds and present improved results with $139\:{\rm fb}^{-1}$ of data \cite{ATLAS:2021bjk}.

\section{Updating mono-\texorpdfstring{$\tau$}{tau} tails at the LHC\label{sec:RES1}}
\subsection{Framework}

We begin by adopting the same effective Lagrangian for NP as in Refs.~\cite{Greljo:2018tzh,Marzocca:2020ueu} to describe the generic transition between up- and down-type quarks:
\begin{align}
\label{eq:lagEFT}
\mathcal{L}&=-\frac{4G_FV_{ij}}{\sqrt{2}}\bigg[g^{ij}_{V_{L}}(\bar{u}_i\gamma_\mu P_L d_j)(\bar{\tau} \gamma^\mu P_L\nu_\tau)\\\nonumber
&+g^{ij}_{V_{R}}(\bar{u}_i\gamma_\mu P_R d_j)(\bar{\tau} \gamma^\mu P_L\nu_\tau)\\\nonumber
&+g^{ij}_{S_L}(\bar{u}_i P_L d_j)(\bar{\tau} P_L\nu_\tau)+g^{ij}_{S_R}(\bar{u}_i P_R d_j)(\bar{\tau} P_L\nu_\tau)\\\nonumber
&+g^{ij}_{T}(\bar{u}_i\sigma_{\mu\nu} P_L d_j)(\bar{\tau} \sigma^{\mu\nu} P_L\nu_\tau)\bigg]+{\rm h.c.},
\end{align}
where $G_F$ is the Fermi constant, $V_{ij}$ are the Cabibbo-Kobayashi-Maskawa (CKM) matrix elements, and $g_X^{ij}$ are the NP couplings, i.e.~Wilson coefficients (WC),  $X\in\{V_L,V_R,S_L,S_R,T\}$. The Lagrangian \eqref{eq:lagEFT} is used to describe $u_i\to d_j\bar{\ell}\nu$ or $d_j\to u_i\ell\bar{\nu}$, but like in previous studies, we also use it to describe $\bar{u}_id_j\to\ell\bar{\nu}$. When considering $\bar{u}_id_j\to\ell\bar{\nu}$ at high energies, we do not integrate out $W$ when describing the Standard Model contribution, which is otherwise considered as background.

Using the above Lagrangian and after neglecting the fermion masses, the partonic cross-section for $b\bar{c}\to \tau\bar{\nu}$ induced by New Physics can be written as:
\begin{align}
\label{eq:partonicCS}
\frac{{\rm d}\hat{\sigma}_{\rm NP}}{{\rm d}\hat{t}} &= \frac{|V_{cb}|^2}{48v^4\pi\hat{s}^2}\bigg[\left(|g_{S_L}|^2+|g_{S_R}|^2\right)\hat{s}^2+4|g_{V_{R}}|^2\hat{u}^2\\\nonumber
&+4|g_{V_{L}}|^2\hat{t}^2+16|g_{T}|^2\left(2\hat{t}^2+2\hat{u}^2-\hat{s}^2\right)\\\nonumber
&+8{\rm Re}(g_T g_{S_L}^*)\left(\hat{u}^2-\hat{t}^2\right)\bigg]\\\nonumber
&+\frac{|V_{cb}|^2}{6v^4}\frac{m_W^2\:\hat{t^2}\:{\rm Re}(g_{V_{LL}})}{\left(\hat{s}-m_W^2\right)\hat{s}^2}\,,
\end{align}
where the electroweak vacuum expectation value is  $v=\left(\sqrt{2}G_F\right)^{-1/2}$, while $s$, $t$ and $u$ are Mandelstam variables. The hats refer to partonic quantities. Only the $g_{V_{L}}$-term may interfere with the SM, which is represented by the last line in \eqref{eq:partonicCS}. After integrating over $\hat{t}\in [-\hat{s},0]$, we get the following partonic cross-section:
\begin{align}
\label{eq:integratedpartonicCS}
\hat{\sigma}(\hat{s})&=\frac{|V_{cb}|^2 \hat{s}}{36v^4\pi}\bigg[|g_{V_{L}}|^2+|g_{V_{R}}|^2+\frac{3}{4}|g_{S_{L}}|^2\\\nonumber
&+\frac{3}{4}|g_{S_{R}}|^2+4|g_T|^2\bigg]+\frac{|V_{cb}|^2}{18v^4}\frac{m_W^2 \:{\rm Re}(g_{V_{L}})\:\hat{s}}{\left(\hat{s}-m_W^2\right)},
\end{align}
where again the last term is the interference between the SM and NP, while  the interference term between the scalar and tensor contributions vanishes in the small fermion mass limit. An important observation at this point is that the NP contribution to the partonic cross-section increases with the center of mass energy with the rate $\propto\hat{s}$. This energy enhancement can partially compensate for the heavy quark PDF suppression at high energies. That is why we focus on the events in the $\text{high-}p_T$ tails.

The hadronic cross-section can be computed by convoluting the partonic cross-section \eqref{eq:integratedpartonicCS} with the parton luminosity functions $\mathcal{L}_{q_i\bar{q}_j}$:
\begin{align}
&\sigma(pp\to \tau^+\nu)=\sum_{ij}\int\frac{dy}{y}\mathcal{L}_{q_i\bar{q}_j}(y)\left[\hat{\sigma}(y s)\right]_{ij},\\
&\mathcal{L}_{q_i\bar{q}_j}=y\int_y^1\frac{dx}{x}\left[f_{q_i}(x,\mu_F)f_{\bar{q}_j}(y/x, \mu_F)+q_i\leftrightarrow \bar{q}_j\right].
\end{align}
$\mathcal{L}_{q_i\bar{q}_j}$ depends on the factorization scale $\mu_F$, which is taken to be the partonic center of mass energy, $\mu_F=\sqrt{\hat{s}}$. The set of PDF's used in this paper are PDF4LHC15\_nnlo\_mc and NNPDF23\_lo\_as\_0130\_qed, provided in the {\tt LHAPDF} package \cite{Buckley:2014ana}.

In practice, we reevaluate the cross-sections during the simulation by using cuts in the phase-space as to match the experimental analysis. However, since the dependence on Wilson coefficients has been made explicit, and since there are no interference terms, we can simply simulate for each $g_X$ separately.

\subsection{Recast}

As mentioned in introduction, we recast the latest ATLAS searches for a heavy resonance decaying to $\tau \nu$, based on $139\:{\rm fb}^{-1}$ of data \cite{ATLAS:2021bjk}. That search is an update of their previous analysis based on $36.1\:{\rm fb}^{-1}$ \cite{ATLAS:2018ihk}. The previous ATLAS result, along with the similar CMS searches \cite{CMS:2018fza}, have been used to derive bounds on the $g_X$ couplings in \cite{Greljo:2018tzh,Marzocca:2020ueu,Iguro:2020keo}.

To simulate the events, we used {\tt MadGraph} version 2.7.3 \cite{Alwall:2014hca}. For each individual contribution to the cross-section we generate $5\times 10^4$ events with up to one jet in the final state. The cross-section for the processes (with and without jets) is computed automatically by {\tt MadGraph} to leading order by using Lagrangian (\ref{eq:lagEFT}), implemented with {\tt Feynrules} \cite{Alloul:2013bka}. The outgoing particles are then showered and hadronized using {\tt Pythia8} \cite{Sjostrand:2014zea}. The ATLAS detector is simulated using {\tt Delphes3} \cite{deFavereau:2013fsa}.

The events are then filtered by imposing the same kinematic cuts as in is the experimental searches. Expressed in terms of the transverse momentum $p_T$, pseudorapidity $\eta$ and the azimuthal angle $\phi$, the cuts are made by selecting events:
\begin{itemize}
\item With one or more reconstructed $\tau$ with ${p_T>30\:{\rm GeV}}$ and ${|\eta|<2.4}$, excluding ${|\eta|\in [1.37,\:1.52]}$\,;\footnote{The previous analysis with $36.1\:{\rm fb}^{-1}$ of luminosity uses a cut of $50\:{\rm GeV}$ instead}
\item With no electron with ${p_T>20\:{\rm GeV}}$ and ${|\eta|<2.47}$, (excluding ${|\eta|\in [1.37,\:1.52]}$), or muon with ${p_T>20\:{\rm GeV}}$ and ${|\eta|<2.5}$\,;
\item For which the transverse missing energy is large, ${E_{\rm miss}>150\:{\rm GeV}}$;
\item Having a reatio between the $\tau$ and neutrino transverse energy within a range ${0.7<p_T/E_{\rm miss}<1.3}$\,;
\item With nearly back to back $\tau$ and neutrino i.e.~${\Delta\phi>2.4}$.
\end{itemize}

The visible cross-section is obtained after binning the remaining events according to the dilepton transverse mass $m_T$,
\begin{align}
m_T=\sqrt{2p_TE_{\rm miss}(1-\cos\Delta\phi)}\,.
\end{align}

The $\tau$ identification efficiency $\epsilon(\tau)$ is taken into account at the generator level, as input for {\tt Delphes3}. It is in general a function of $p_T(\tau)$, but is reported by ATLAS as a function of $m_T$ \cite{ATLAS:2018ihk}. In the CMS study it is instead reported as a constant value, independent of $p_T(\tau)$, but with an error that grows linearly with $p_T(\tau)$. In this Section we only consider the central value given for this efficiency, without accounting for the error, just like it was done in Refs.~\cite{Greljo:2018tzh,Marzocca:2020ueu}. The effect of including the systematics will be explored in Section \ref{sec:eff}.

We validate the recast by comparing the acceptance ($\mathcal{A}$) and efficiency $(\epsilon)$ with results by ATLAS with $36\:{\rm fb}^{-1}$ and with $139\:{\rm fb}^{-1}$, respectively, on a $W'$ toy model. Indeed, we find that our final values for $\mathcal{A}\times\epsilon$ agree, within $10~\%$ with the result by ATLAS.

\subsection{Analysis}

We consider the observed number of events, simulated background, and background uncertainty as reported in Fig. 5 of Ref.~\cite{ATLAS:2021bjk}. We take the background and its uncertainty from ATLAS and consider the different uncertainties to be completely uncorrelated. To obtain the bounds on the Wilson coefficients, we use the package {\tt pyhf} \cite{Heinrich:2021gyp} in which the CLs method has been implemented~\cite{Cowan:2010js} in order to compute the 95$~\%$ limit.

We should emphasize that using the Confidence Levels (CLs) method, we avoid the problem of bins with a small number of events, that could otherwise be problematic if we used a naive $\chi^2$-analysis. This problem is particularly pronounced in our case because the signal-to-background ratio increases with energy, but the background rapidly approaches zero. Moreover, the most significant part of the spectrum for our purpose is the one with the least statistics. To circumvent this problem and compare the two methods, we tried merging bins in order to justify the Gaussian approximation needed in the $\chi^2$-approach. That way we find that the results of CLs methods and $\chi^2$-approach are indeed compatible.

\subsection{Results}
We set limits on individual Wilson coefficients $g_X$, with $X\in \{V_L,V_R,S_L,S_R,T\}$ by solving for ${{\rm CL}_{\rm obs}}(g_i)>0.95$. We also include the scenarios in which $g_{S_L}=\pm 4g_T$ at $\mu\simeq 1\:{\rm TeV}$, which can occur in peculiar leptoquark scenarios. The resulting limits are summarized in Tab~\ref{tab:WCLimitsFromDirectSearches}. The results are reported at $\mu=m_b$, as obtained using the renormalization group running with anomalous dimensions given in Ref.~\cite{Becirevic:2020rzi}, which also changes the ratio between $g_{S_L}$ and $g_T$ in the last two lines.
\begin{table}[h]
\renewcommand{\arraystretch}{1.5}
\centering
\begin{tabular}{c|ccc|c|}
\cline{2-5}
\multicolumn{1}{c|}{} & \multicolumn{3}{c|}{ATLAS} &\multicolumn{1}{c|}{CMS } \\ \hline
\multicolumn{1}{|c|}{$\mathcal{L}\:\:\:({\rm fb}^{-1})$} & 36.1 & 139 & \color{gray}3000 & 35.9 \\ \hline\hline
\multicolumn{1}{|c|}{$|g_{V_L}|, |g_{V_R}|$} & 0.57 & 0.32 & \color{gray}0.15 & 0.32  \\
\multicolumn{1}{|c|}{$|g_{S_L}|, |g_{S_R}|$} & 1.13 & 0.57 & \color{gray}0.33 & 0.52\\
\multicolumn{1}{|c|}{$|g_{T}|$} & 0.32 & 0.18 & \color{gray}0.08 & 0.17 \\
\multicolumn{1}{|c|}{$|g_{S_L}|=8.1|g_{T}|$} & 1.00 & 0.51 & \color{gray}0.27 & 0.48\\
\multicolumn{1}{|c|}{$|g_{S_L}|=8.5|g_{T}|$} & 1.08 & 0.56 & \color{gray}0.29 & 0.52\\ \hline
\end{tabular}
\caption{Upper limits on the individual Wilson coefficients at $\mu=m_b$ using the ATLAS data with $36$ and $139\:{\rm fb}^{-1}$, respectively~\cite{ATLAS:2018ihk,ATLAS:2021bjk} and neglecting the systematic uncertainty on the $\tau$ reconstruction efficiency, as done in Refs.~\cite{Greljo:2018tzh,Marzocca:2020ueu}. Projection to $3~{\rm ab}^{-1}$ is also quoted, using the method described in the text. The above values are obtained by using only the central value of the $\tau$ reconstruction efficiency, $\epsilon(\tau)$. The last two lines correspond respectively to $g_{S_L}=+4g_T$ and $g_{S_L}=-4g_T$ at a scale of $\mu\simeq 1\:{\rm TeV}$.}
\label{tab:WCLimitsFromDirectSearches}
\end{table}

In order to project the current limits to the targeted $3\:{\rm ab}^{-1}$ of data, we proceed as follows:
\begin{itemize}
    \item We multiply the signal and background by the increase in luminosity.
    \item We multiply the background uncertainty by the square root of the increase in luminosity.
    \item We use the expected CL instead of the observed one.
\end{itemize} 
If all the errors were Gaussian, by going from luminosity $\mathcal{L}_0$ to $\mathcal{L}_1$  we would expect an improvement in the limit proportional to $\left(\mathcal{L}_1/\mathcal{L}_0\right)^{1/4}$. In this way one expects the reduction of the limit by $30~\%$ when going from $36.1$ to $139\:{\rm fb}^{-1}$. According to Tab.~\ref{tab:WCLimitsFromDirectSearches}, the actual data exhibit even better scaling, sitting at around $45~\%$. This makes us confident that by following the same scaling, the projection to $3\:{\rm ab}^{-1}$ is reasonable and would improve (reduce) the limit by about $50~\%$.

\section{Impact of the \texorpdfstring{$\tau$}{tau} reconstruction efficiency\label{sec:RES2}}
\label{sec:eff}

From the results shown in Tab.~\ref{tab:WCLimitsFromDirectSearches} we observe a sizeable difference in the limits obtained from CMS and ATLAS data. Such a difference was already spotted in Ref.~\cite{Greljo:2018tzh}. Considering the similarities between the two detectors, in particular the fact that the $\tau$ leptons are treated in a similar manner in both searches, it is surprising to have such a large gap in sensitivity, equivalent to a fourfold increase in luminosity.

We investigate whether or not it is possible to understand this difference from the treatment of systematic errors on the $\tau$ reconstruction efficiency. 

In Ref.~\cite{ATLAS:2018ihk}, the efficiency reported by ATLAS is a function of $m_T$, namely:\\
{\footnotesize
``The reconstruction efficiency depends on $m_T$, $\epsilon(m_T {\rm\:[TeV]})=0.633-0.313m_T+0.0688m_T^2-0.00575m_T^3$. [...] The relative uncertainty in the parameterized efficiency due to the choice of signal model is $\sim 10~\%$. [...]. An additional uncertainty that increases by 20-25~\% per TeV is assigned to $\tau_{\rm had-vis}$ candidates with $p_T > 150$."}

Instead, the efficiency in the CMS study \cite{CMS:2018fza} is discussed as follows:\\
{\footnotesize
``This working point has an efficiency of about 70~\% for genuine $\tau_h$ [...].  The uncertainty associated with the $\tau_h$ identification is 5~\% [48]. An additional systematic uncertainty, which dominates for high-$p_T$ $\tau_h$ candidates, is related to the degree of confidence that the MC simulation correctly models the identification efficiency. This additional uncertainty increases linearly with $p_T^\tau$ and amounts to $+5~\%/-35~\%$ at $p_T^\tau =1\:{\rm TeV}$."}

\begin{figure}[H]
\centering
\includegraphics[width=9cm]{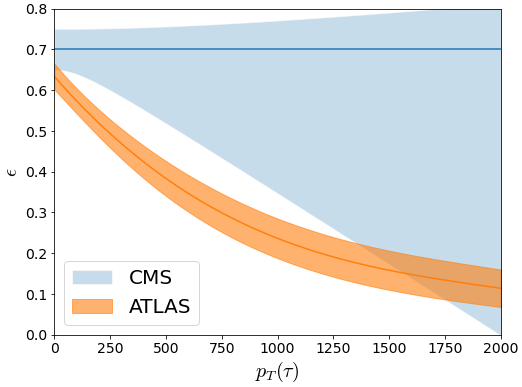}
\caption{Comparison between the ATLAS and CMS $\tau$ reconstruction efficiency as reported in Refs \cite{ATLAS:2018ihk} and \cite{CMS:2018fza} respectively. In this plot we made the simplified assumption that $m_T\simeq 2p_T$, which is true as long as the neutrino is the only missing energy of the event and no other jet is found.}
\label{fig:ptEffError}
\end{figure}

In Fig. \ref{fig:ptEffError} we plot the two efficiencies~\cite{ATLAS:2018ihk,CMS:2018fza}.  Note that in Ref.~\cite{Marzocca:2020ueu} the value $\epsilon(\tau) = 0.7$ has been used, as extracted from the Fig.~4 of Ref.~\cite{CMS:2018jrd} in which only the interval in $p_T^\tau$ between $20~\mathrm{GeV}$ and $100~\mathrm{GeV}$ is considered. Instead of giving a flat value for $\epsilon(\tau)$, with asymmetric uncertainties, it would be very helpful if CMS also provided a function to describe the downward trend of the $\tau$ reconstruction efficiency.

Since the SM cross-section decreases faster than the NP contribution, the last few bins in the CLs method are always the most important ones. In the case of CMS, neglecting the uncertainty on the $\tau$ reconstruction efficiency in the last bins, where it is the largest and where the events contribute the most to the final $p$-value, cf. Fig.~\ref{fig:ptEffError}, can lead to a too stringent constraint on the Wilson coefficient.

To take into account the effect of this systematic error we add a posterior reconstruction efficiency with its uncertainty, reported in the experimental papers. In the CLs method this can be done by assigning the signal a correlated error. 
To illustrate the effect of inclusion of that systematics we again recast the CMS and ATLAS searches from the $36~\mathrm{fb}^{-1}$ datasets and obtain the results presented in Tab.~\ref{tab:WCLimitsFromDirectSearcheswithEff}. Note also that the same procedure is used with new ATLAS data.

From a comparison between the results given in Tab.~\ref{tab:WCLimitsFromDirectSearches} and Tab.~\ref{tab:WCLimitsFromDirectSearcheswithEff} we see that the bounds become weaker by $\sim 10~\%$ for ATLAS and by $\sim 40~\%$ for CMS, after the systematic error on the $\tau$ reconstruction efficiency is included.

\begin{table}[H]
\renewcommand{\arraystretch}{1.5}
\centering
\begin{tabular}{c|ccc|c|}
\cline{2-5}
\multicolumn{1}{c|}{} & \multicolumn{3}{c|}{ATLAS} &\multicolumn{1}{c|}{CMS } \\ \hline
\multicolumn{1}{|c|}{$\mathcal{L}\:\:\:({\rm fb}^{-1})$} & 36.1 & 139 & \color{gray}3000 & 35.9 \\ \hline
\multicolumn{1}{|c|}{$|g_{V_L}|, |g_{V_R}|$} & 0.69 & 0.36 & \color{gray}0.17 & 0.44 \\
\multicolumn{1}{|c|}{$|g_{S_L}|, |g_{S_R}|$} & 1.29 & 0.62 & \color{gray}0.32 & 0.75\\
\multicolumn{1}{|c|}{$|g_{T}|$} & 0.34 & 0.19 & \color{gray}0.09 & 0.28\\
\multicolumn{1}{|c|}{$|g_{S_L}|=8.1|g_{T}|$} & 1.14 & 0.55 & \color{gray}0.29 & 0.70\\
\multicolumn{1}{|c|}{$|g_{S_L}|=8.5|g_{T}|$} & 1.23 & 0.60 & \color{gray}0.31 & 0.76\\ \hline
\end{tabular}
\caption{Same as in Tab.~\ref{tab:WCLimitsFromDirectSearches} but with the inclusion of systematic error on the $\tau$ reconstruction efficiency. See Sec.~\ref{sec:eff} for details.}
\label{tab:WCLimitsFromDirectSearcheswithEff}
\end{table}

\section{Effects of the mediator propagation}
\label{sec:eft}
Due to the energy enhancement in the partonic cross-section \eqref{eq:integratedpartonicCS} the most significant events in the analysis will be located at the high energy end of the spectrum, i.e. in the tail of the distribution. For obtaining the results presented in Sections~\ref{sec:RES1} and \ref{sec:RES2}, as well as in Refs.~\cite{Greljo:2018tzh,Marzocca:2020ueu,Iguro:2020keo}, the most constraining events are indeed those above $1\: {\rm TeV}$.~\footnote{Note that in the new ATLAS data \cite{ATLAS:2021bjk} $60$ events are found above $1\: {\rm TeV}$.} If we assume NP to be mediated by a $t$- or $u\text{-channel}$ particle with a mass around  $1\:{\rm TeV}$, the partonic cross-section will be greatly overestimated if we do not account for the propagation of the NP state. This can be easily seen if we write the propagator
\begin{align}
\frac{1}{t-m^2}&\simeq-\frac{1}{m^2} \left( 1 + \frac{t}{m^2} +\dots \right), \qquad t\in [-s, 0].
\label{eq:order1}
\end{align}
Obviously, if the negative $t/m^2$ correction is neglected the cross-section would be much larger than the one obtained by using the usual EFT approach. This effect becomes even worse due to the energy enhancement in the cross-section and the fact that the size of the $t/m^2$ correction is actually enhanced by the cuts in the analysis.

To illustrate the effect of propagation of the mediator we shall compare the recast results of a leptoquark (LQ) model with and without the use of EFT.

\subsection{Framework}

As an example, we focus on the $R_2$ leptoquark Lagrangian \cite{Dorsner:2016wpm},
\begin{align}
\mathcal{L}_{R_2}&=y_R^{ij}\bar{Q}_iR_2l_{Rj}+y_L^{i,j}\bar{u}_{Ri}\widetilde{R}_2^\dagger L_j+h.c.
\label{eq:R2Lag}
\end{align}
where, as usual, $Q$ and $L$ are the left-handed doublets of quarks and leptons, while $u_R$ and $l_R$ are the right-handed singlets. We work in the basis where down-quark Yukawa couplings are diagonal. In Eq.~\eqref{eq:R2Lag} we use the notation with $\widetilde{R}_2=i\tau_2 R_2^*$, where $\tau_2$ is the usual Pauli matrix. As for the Yukawa couplings, i.e. between the $R_2$ leptoquark and the lepton and quark flavors, we use:
\begin{align}
y_L=\begin{pmatrix}
0&0&0\\
0&0&y_L^{c\tau}\\
0&0&0
\end{pmatrix},\qquad
y_R=\begin{pmatrix}
0&0&0\\
0&0&0\\
0&0&y_R^{b\tau}
\end{pmatrix}.
\label{eq:yukdef}
\end{align}
This is the minimal set of couplings needed to explain the charged current $B$-anomalies~\cite{Becirevic:2018afm,Angelescu:2021lln}, see also Ref.~\cite{Iguro:2020keo}. We take the mass $m_{R_2}$ as our benchmark point, which is consistent with current limits derived from direct searches~\cite{Becirevic:2018afm,Angelescu:2021lln}. After neglecting the fermion masses, the partonic cross-section reads
\begin{align}
\label{eq:R2CS}
\frac{{\rm d}\hat{\sigma}\left(c\bar{b}\to\tau^+\nu_\mu\right)}{{\rm d}\hat{t}} &= \frac{1}{64N_c\pi \hat{s}^2}\frac{\left|y^L_{c\tau}\right|^2\left|y^R_{b\tau}\right|^2\hat{u}^2}{\left(\hat{u}-m_{R_2}^2\right)^2}\, .
\end{align}
This can be easily integrated since in the same limit $\hat{s}+\hat{t}+\hat{u}=0$. We get
\begin{align}
\hat{\sigma}(\hat{s})&\simeq\frac{|y^L_{b\tau}|^2|y^R_{c\tau}|^2}{192\pi m_{R_2}^2}\left[\frac{x+2}{x(1+x)}-\frac{2\log(1+x)}{x^2}\right],
\label{eq:FullPartonic}
\end{align}
where $x=\hat{s}/m_{R_2}^2$.

We recast the latest ATLAS searches \cite{ATLAS:2021bjk}, using the same selection of events as in the previous Section, again by allowing at most one extra jet in the final state.

\subsection{Results}

Like in the previous Section, we use the CLs method to constraint the two non-zero couplings appearing in Eq.~(\ref{eq:R2Lag},~\ref{eq:yukdef}). The resulting limit can be expressed as
\begin{align}
\boxed{
\left|y^L_{c\tau}\right|\left|y^R_{b\tau}\right|<2.44 } \qquad \text{for}\quad m_{R_2}=1.3\:{\rm TeV}\,.
\end{align}
This inequality translates to a bound on the Wilson coefficient $g_{S_L}\equiv g_{S_L}(m_b)$ as
\begin{align*}
|g_{S_L}|=8.1\:|g_T|\ <\ 0.88\, ,
\end{align*}
which is to be compared with $|g_{S_L}|<0.55$ as given in Tab.~\ref{tab:WCLimitsFromDirectSearcheswithEff}.

\begin{figure}[H]
\centering
\includegraphics[width=8cm]{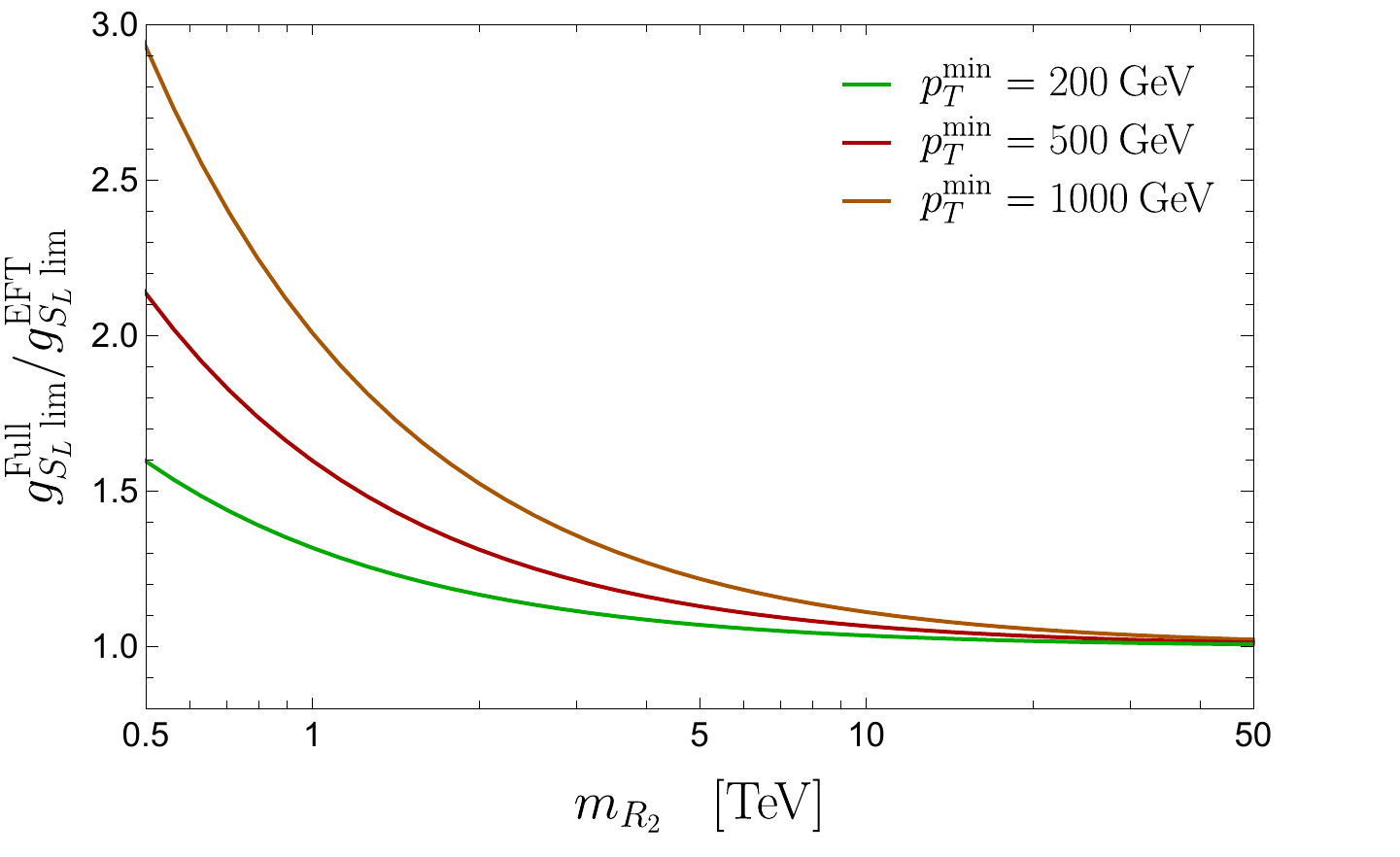}
\caption{Ratio of the limits obtained with the full model and the EFT for various $p_T$ cuts, in the limit of infinite luminosity.}
\label{fig:limitgsl}
\end{figure}

We see that the latter constraint on $g_{S_L}$ is approximately $60~\%$ weaker than the one obtained by using EFT alone. This statement is obviously dependent on the leptoquark mass and the difference should decrease as $m_{R_2}$ increases (see Fig.~\ref{fig:limitgsl}).

\begin{figure}[H]
\centering
\includegraphics[width=8cm]{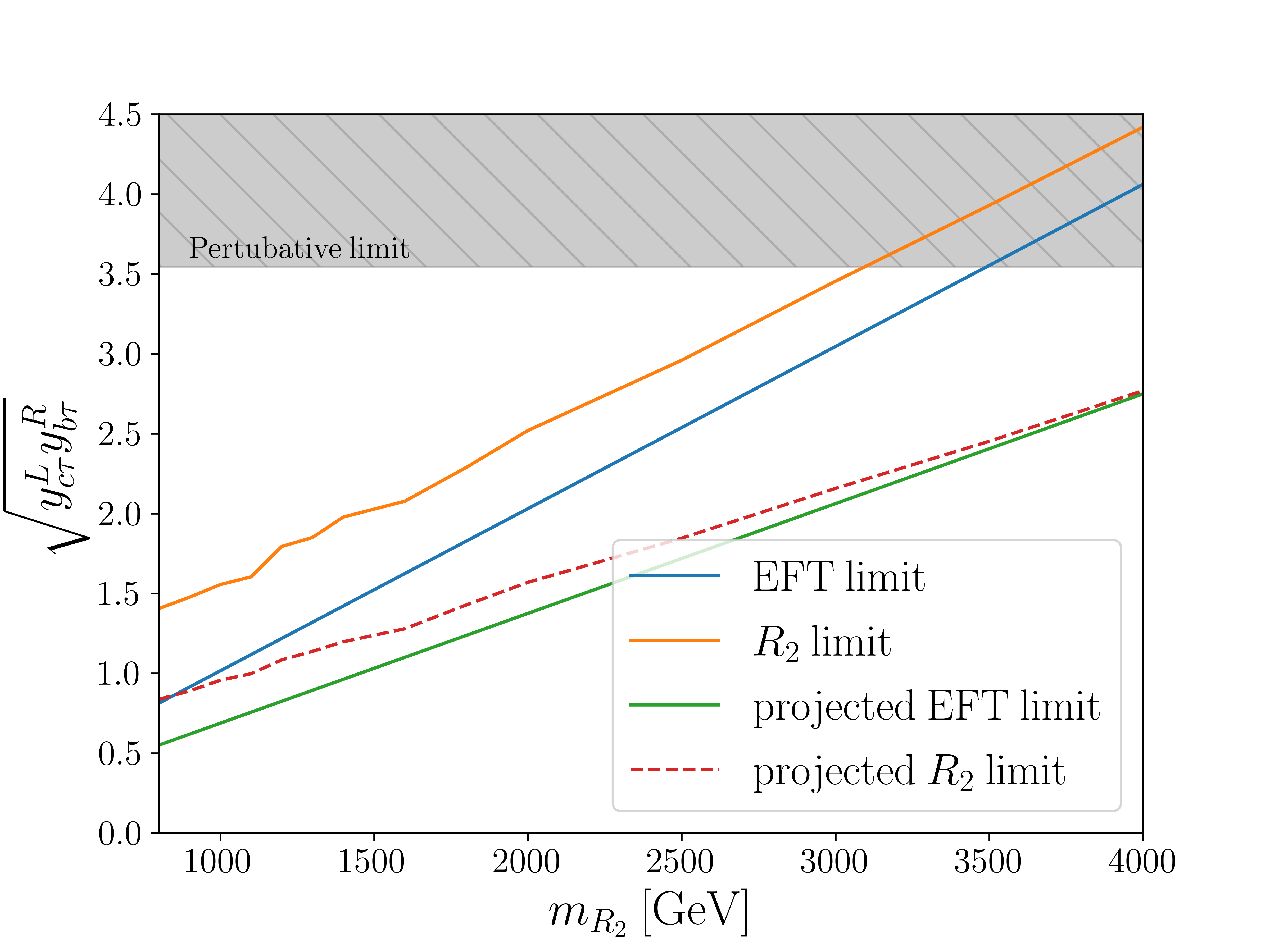}
\caption{$95~\%$ exclusion limit on $\sqrt{y^L_{c\tau}y^R_{b\tau}}$  as a function of the $R_2$ mass (orange). The equivalent EFT result is shown in blue. We also include the projected limits with $3~{\rm ab}^{-1}$ of integrated luminosity.}
\label{fig:ylim}
\end{figure}

An important observation is that despite the total cross-section being only $10~\%$ smaller when considering the propagation of the mediator the difference in the limits on Wilson coefficients can be much larger. This is because the ratio of cross-sections depends on $p_T^{\rm min}$, the value starting from which the high-$p_T$ tail is defined in order to derive desired constraints. For a smaller $p_T^{\rm min}$, the $t/m^2$ term in Eq.~\eqref{eq:order1} is large only in a small part of the phase space, whereas for larger values of $p_T^{\rm min}$ this term remains always large. In Fig.~\ref{fig:limitgsl} we show how the two constraints compare for various $p_T^{\rm min}$ and various values of leptoquark masses. Indeed for heavier leptoquarks the constraints on couplings derived from EFT and propagating leptoquarks become closer to each other. Already at $m_{R_2}\simeq 5\:{\rm TeV}$, we find that the two constraints are within $20~\%$ from each other. One should keep in mind, however, that the bound on $|g_{S_L}|$ is stronger for larger $p_T^{\rm min}$ because the events larger than $1\:{\rm TeV}$ provide the most significant constraints.

in Fig.~\ref{fig:ylim} we plot the limit on the average coupling $\sqrt{y^L_{c\tau}y^R_{b\tau}}$, that we obtain from the ATLAS data. We see that we can only probe the region under $3\:{\rm TeV}$ because the couplings would otherwise become nonperturbative, i.e.~larger than $\sim\sqrt{4\pi}$. However, with a further increase in luminosity, the slope of the curves shown in Fig.~\ref{fig:ylim} would become smaller, thus opening a possibility to probe heavier leptoquarks. We also tried to include the effect of the leptoquark width in this analysis, and we found that the effect is insignificant in the perturbative regime,  $|y^L_{c\tau}|,|y^R_{b\tau}|<\sqrt{4\pi}$.

\section{CONCLUSIONS and DISCUSSION}

In this letter, we provided a reinterpretation of the latest $139\:{\rm fb}^{-1}$ mono-$\tau$ search at LHC \cite{ATLAS:2021bjk} in order to provide bounds of the New Physics couplings relevant to the semileptonic dimension-6 operators. We found that the new ATLAS limits are about a factor of $2$ stronger than those obtained in the previous analyses, with $36\:{\rm fb}^{-1}$. 

By proceeding like in the previous theoretical recasts, we also find a puzzling difference between the bounds on the Wilson coefficients as derived from the ATLAS and from the CMS data, despite their same luminosity of $36~\mathrm{fb}^{-1}$. We show that this difference can be partly explained by the inclusion of the systematic error on the $\tau$ lepton reconstruction efficiency, which was not accounted for in the previous theory analyses. We find that the inclusion of that systematics would loosen the bounds on Wilson coefficients by $10~\%$ and $40~\%$ when using the ATLAS and CMS data respectively. 

Our final results are reported in Tab.~\ref{tab:WCLimitsFromDirectSearcheswithEff}.

Finally, we examine the difference between the bounds on Wilson coefficients obtained by including in the analysis the propagation of the NP particle of mass $\mathcal{O}(1\,\mathrm{TeV})$, with those obtained by using the EFT approach alone, i.e. by integrating out the NP particle. 
The nonresonant nature of the process does not exempt it from the separation of scale requirement, necessary for validity of the EFT description. 
We find that the inclusion of propagation further relaxes the bounds on Wilson coefficients. In particular, such a bound for the case of the $R_2$ leptoquark becomes $60~\%$ larger when the propagation of the $m_{R_2}=1.3$~TeV leptoquark is included in the analysis. Therefore, the inclusion of propagation in the analyses in phenomenological studies is important.
\vspace*{0.8cm}
\section*{ACKNOWLEDGEMENTS}

We thank D.~Becirevic, D.~Faroughy, O.~Sumensari and F.~Wilsch for their continuous help and encouragement during this project. This work has been supported in part by the European Union's Horizon 2020 research and innovation programme under the Marie Sklodowska-Curie grant agreements N$^\circ$~674896 and 690575.

%\bibliographystyle{hplain}
%\bibliography{references}

\end{document}